\documentclass[floatfix,aps,superscriptaddress,prd,amsmath,nofootinbib,preprintnumbers,onecolumn]{revtex4}

\usepackage{graphicx}
\usepackage{bm}
\usepackage{float}
\usepackage[
colorlinks=true,
citecolor=blue,
linkcolor=blue,
urlcolor=blue
]{hyperref}

\newcommand{\nc}{\newcommand*}
\nc{\Om}{\Omega}
\nc{\ogw}{\Omega_{\mathrm{GW}}}
\nc{\rd}{\mathrm{d}}
\nc{\eg}{\textit{e.g.~}}
\nc{\red}[1]{\textcolor{red}{#1}}
\nc{\lvc}{LIGO/Virgo}

\def\half{{1\over 2}}

\def\half{{1\over 2}}
\def\({\left(}
\def\){\right)}
\def\[{\left[}
\def\]{\right]}
\def\e{\begin{equation}}
\def\q{\end{equation}}
\def\m{\begin{eqnarray}}
\def\n{\end{eqnarray}}

\begin{document}

\title{Constraints on the canonical single-field slow-roll inflation model from observations}

\author{Jun~Li}
\affiliation{Qingdao Key Laboratory of Novel Optoelectronic Devices and Ultrafast Intelligent Manufacturing, School of Mathematics and Physics, Qingdao University of Science and Technology, Qingdao 266061, China.}
\affiliation{CAS Key Laboratory of Theoretical Physics, Institute of Theoretical Physics, Chinese Academy of Sciences, Beijing 100190, China}

\thanks{This work is accepted for publication in \emph{Chinese Physics C}.}

\author{Guanghai Guo}
\affiliation{Qingdao Key Laboratory of Novel Optoelectronic Devices and Ultrafast Intelligent Manufacturing, School of Mathematics and Physics, Qingdao University of Science and Technology, Qingdao 266061, China.}

\author{Pengfei Yan}
\affiliation{Qingdao Key Laboratory of Novel Optoelectronic Devices and Ultrafast Intelligent Manufacturing, School of Mathematics and Physics, Qingdao University of Science and Technology, Qingdao 266061, China.}

\author{Xiong Yang}
\affiliation{College of Science, Civil Aviation University of China, Tianjin 300300, China.}

\date{\today}

\begin{abstract}
In this work, we employ two complementary approaches to constrain the canonical single-field slow-roll inflation scenario. The first method makes explicit use of the analytic dependence of the primordial perturbations on the slow-roll parameters, whereas the second adopts a phenomenological parameterization of the primordial scalar and tensor spectra. Using the latest observational datasets, including Planck satellite data, BICEP/Keck measurements, baryon acoustic oscillation data, and the recent DESI Data Release 2, we derive direct constraints on the slow-roll parameters. A key advantage of this strategy is that it allows us to compute the predictions of single-field slow-roll inflation directly from the constrained parameter values. We illustrate the resulting predictions for the parameters that characterize the scalar power spectrum and place constraints on several representative inflationary models. Our analysis shows that monomial-potential inflation is disfavored, while models with concave potentials, such as the Starobinsky model and brane inflation, are preferred. From the constraints on the slow-roll parameters, we obtain a tensor spectral index in the single-field slow-roll framework that is very small, $|n_t|\lesssim 4.9\times 10^{-3}$, a value that will be challenging to measure with CMB data alone in the foreseeable future. Moreover, the absolute value of the derived running of the tensor spectral index does not exceed $1.91\times 10^{-4}$ at $95\%$ confidence level, based on the combination of CMB+BAO+DESI datasets.

\end{abstract}

\maketitle

\section{introduction}
Inflation is broadly regarded as the leading paradigm for describing the physics of the very early Universe \cite{Guth:1980zm,Starobinsky:1980te,Linde:1981mu,Albrecht:1982wi}. It naturally resolves several longstanding puzzles of the standard hot big bang model, most notably the flatness, horizon, and monopole problems, by postulating a phase of near-exponential expansion that renders the observable Universe highly homogeneous and isotropic. More importantly, inflation provides a quantum-mechanical origin for the primordial density perturbations that later seed the temperature anisotropies observed in the cosmic microwave background (CMB) and initiate the formation of cosmic structure.

In addition to scalar perturbations, inflation generically excites tensor perturbations, i.e., primordial gravitational waves, which constitute a fundamental degree of freedom of the gravitational field. These tensor modes imprint a unique signature on the polarization pattern of the CMB, the so-called primordial B-mode signal \cite{Li:2017cds,Li:2018iwg,Li:2019vlb,Li:2021scb,Li:2024cmk,Li:2021nqa}. Beyond the linear tensor perturbations produced directly by the inflationary background, second-order scalar perturbations can also source a stochastic gravitational-wave background, often referred to as scalar-induced gravitational waves \cite{Li:2022avp,Li:2021uvn,Li:2023uhu,Chen:2024fir}. The detection of either type of gravitational-wave signal would open a new observational window onto the energy scale and dynamics of inflation.

During the inflationary phase, the Hubble parameter remains approximately constant, leading to a nearly scale-invariant spectrum of curvature perturbations. This characteristic prediction arises because quantum fluctuations of the inflaton field are stretched to superhorizon scales and subsequently freeze, providing the initial conditions for structure formation. The simplest class of inflationary models, namely canonical single-field slow-roll scenarios, makes a set of correlated predictions for the scalar and tensor power spectra, offering a clear target for observational tests. Current and future CMB experiments, large-scale structure surveys, and gravitational-wave observatories are therefore poised to test the inflationary paradigm with unprecedented precision, potentially distinguishing among competing models.

Since inflation occurred in the first moments of the Universe, its imprints are encoded in the statistical properties of cosmic structures. The most precise probes to date come from observations of the cosmic microwave background (CMB), especially its temperature anisotropies and polarization patterns. Satellite missions such as Planck have provided high-fidelity, full-sky maps of the CMB, enabling detailed constraints on the amplitude and spectral shape of primordial curvature perturbations \cite{Planck:2018vyg}. Ground-based experiments, most notably the BICEP/Keck array with data through the 2018 observing season, have further tightened limits on the tensor-to-scalar ratio by searching for the distinctive B-mode polarization signal of primordial gravitational waves \cite{BICEP:2021xfz}.

To break degeneracies among cosmological parameters and sharpen inferences about the primordial power spectra, we complement CMB measurements with large-scale structure tracers. In this work, we include baryon acoustic oscillation (BAO) data from galaxy surveys \cite{Beutler:2011hx,Ross:2014qpa,BOSS:2016wmc}, notably the latest DESI Data Release 2 \cite{DESI:2025zgx}. The addition of BAO distances improves constraints on the background expansion history and, through the integrated Sachs-Wolfe and redshift-space distortion effects, helps to separate the primordial signal from late-time astrophysical processes.

In this paper, we replace the conventional phenomenological parametrization of the primordial power spectra with slow-roll parameters in a CosmoMC analysis. We derive the corresponding scalar and tensor spectral parameters and compare them with several representative inflationary models. The novelty and motivation of this work are twofold: first, we test the theoretical relations between the power-spectrum observables and the slow-roll parameters; second, we present a direct comparison between constraints derived from the slow-roll framework and those obtained from a phenomenological sampling approach.

We adopt two complementary methodologies to confront inflationary predictions with the current ensemble of cosmological and gravitational-wave data. The first approach exploits the analytic slow-roll framework, in which the scalar and tensor power spectra are expressed directly in terms of the slow-roll parameters. This method retains the full theoretical linkage imposed by single-field slow-roll dynamics, allowing the data to constrain the inflationary potential in a model-independent manner. The second approach employs a phenomenological parameterization of the primordial spectra, treating the spectral indices, their running, and the tensor-to-scalar ratio as free parameters. This model-agnostic strategy makes minimal theoretical assumptions and serves as a robustness check on the constraints derived from the slow-roll description. By comparing the outcomes of the two approaches, we can assess whether the current data exhibit a preference for the specific correlation patterns predicted by canonical slow-roll inflation or whether they allow for more general early-Universe scenarios.

\section{the canonical single-field slow-roll inflation model}
In this section, we focus on the canonical single-field slow-roll inflation model, in which inflation is driven by the inflaton potential $V(\phi)$. The inflationary dynamics are governed by
\m
&&H^2={1\over 3 M_p^2}\[\half {\dot\phi}^2+V(\phi)\], \\
&&\ddot \phi+3H\dot\phi+V^\prime (\phi)=0,
\n
where $M_p=1/\sqrt{8\pi G}$ is the reduced Planck energy scale, and the dot and prime denote derivatives with respect to the cosmic time $t$ and the inflation field $\phi$, respectively. The inflation field slowly rolls down its potential if $\epsilon\ll 1$ and $|\eta|\ll 1$, where
\m
\epsilon&=&\frac{M_p^2}{2}\(\frac{V^\prime\(\phi\)}{V\(\phi\)}\)^2,\\
\eta&=&M_p^2\frac{V^{\prime\prime}\(\phi\)}{V\(\phi\)}.
\n
The power spectra of the scalar and tensor perturbations are given in \cite{Huang:2014yaa} by
\m
P_s&\simeq& \[1+{25-9c\over 6}\epsilon-{13-3c\over 6}\eta\] {V/M_p^4\over 24\pi^2 \epsilon}, \\
P_t&\simeq& \[1-{1+3c\over 6}\epsilon\]{V/M_p^4\over 3\pi^2/2},
\n
where $c\simeq 0.08145$. The power spectra of scalar and tensor perturbations are parameterized as
\m
P_s(k)&=&A_s\(\frac{k}{k_*}\)^{n_s-1+\frac{1}{2}\alpha_s\ln(k/k_*)+\frac{1}{6}\beta_s(\ln(k/k_*))^2+...}, \label{eqs:spectrumscalar}\\
P_t(k)&=&A_t\(\frac{k}{k_*}\)^{n_t+\frac{1}{2}\alpha_t\ln(k/k_*)+...},\label{eqs:spectrumtensor}
\n
where $A_s(A_t)$ denotes the scalar (tensor) amplitude at the pivot scale $k_*=0.05$ Mpc$^{-1}$, $n_s$ is the scalar spectral index, $\alpha_s\equiv\mathrm{d} n_s/\mathrm{d}\ln k$ is the running of the scalar spectral index, $\beta_s\equiv{\mathrm{d}^2n_s}/{\mathrm{d}\ln k^2}$ is the running of the running of the scalar spectral index, $n_t$ is the tensor spectral index, and $\alpha_t\equiv \mathrm{d} n_t/\mathrm{d}\ln k$ is the running of the tensor spectral index. It is customary to introduce a new parameter, namely the tensor-to-scalar ratio $r$, to quantify the tensor amplitude relative to the scalar amplitude at the pivot scale:
\e
r\equiv\frac{A_t}{A_s}.
\q
The relations between the power-spectrum parameters and the slow-roll parameters are given in \cite{Huang:2014yaa,Gong:2001he,Casadio:2004ru,Lorenz:2008et,Martin:2013uma,Auclair:2026qtk,Auclair:2022yxs}
\m
r&\approx&16\epsilon\[1-\frac{13-3c}{6}\(2\epsilon-\eta\)\],\label{slr}\\
n_t&\approx&-2\epsilon-\frac{2\(2+3c\)}{3}\epsilon^2-\frac{1-3c}{3}\epsilon\eta,\label{slnt}\\
\alpha_t&\approx&-8\epsilon^2+4\epsilon\eta-\frac{8\(5+6c\)}{3}\epsilon^3+2\(1+7c\)\epsilon^2\eta+2\(1-c\)\epsilon\eta^2,\label{slalphat}\\
n_s&\approx&1-6\epsilon+2\eta+\frac{2(22-9c)}{3}\epsilon^2-2(7-2c)\epsilon\eta+\frac{2}{3}\eta^2,\label{slns}\\
\alpha_s&\approx&-24\epsilon^2+16\epsilon\eta-2\xi+\frac{8\(41-18c\)}{3}\epsilon^3-\frac{4\(109-36c\)}{3}\epsilon^2\eta+4\(9-2c\)\epsilon\eta^2+2\(11-3c\)\epsilon\xi-\frac{25-3c}{6}\eta\xi,\label{slalphas}\\
\beta_s&\approx&-192\epsilon^3+192\epsilon^2\eta-32\epsilon\eta^2-24\epsilon\xi+2\eta\xi+2\sigma+96\(13-6c\)\epsilon^4
-\frac{8\(791-288c\)}{3}\epsilon^3\eta+\frac{16\(173-48c\)}{3}\epsilon^2\eta^2\label{slbetas}\\&&\nonumber-\frac{8\(31-6c\)}{3}\epsilon\eta^3+\frac{4\(235-72c\)}{3}\epsilon^2\xi-\frac{511-111c}{3}\epsilon\eta\xi
+\frac{29-3c}{6}\eta^2\xi+\frac{25-3c}{6}\xi^2-\frac{103-27c}{3}\epsilon\sigma+\frac{55-9c}{6}\eta\sigma,
\n
where
\m
\xi&=&M_p^4\frac{V^\prime\(\phi\)V^{\prime\prime\prime}\(\phi\)}{V^2\(\phi\)},\\
\sigma&=&M_p^6\frac{{V^\prime}^2\(\phi\)V^{\prime\prime\prime\prime}\(\phi\)}{V^3\(\phi\)}.
\n
See, for example, related references in \cite{Stewart:1993bc,Leach:2002ar,Huang:2006yt}.

\section{constraints on the slow-roll parameters from observations}
\begin{figure}
\centering
\includegraphics[width=9cm]{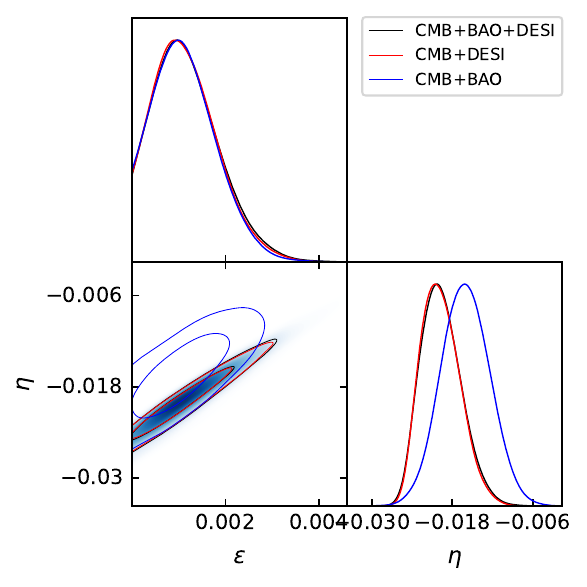}
\caption{The contour plots and likelihood distributions of the slow-roll parameters $\{\epsilon,\eta\}$ are shown at the $68\%$ and $95\%$ confidence levels, derived from the CMB+BAO+DESI, CMB+DESI, and CMB+BAO data combinations, respectively.}
\label{figure1}
\end{figure}

\begin{figure}
\centering
\includegraphics[width=11cm]{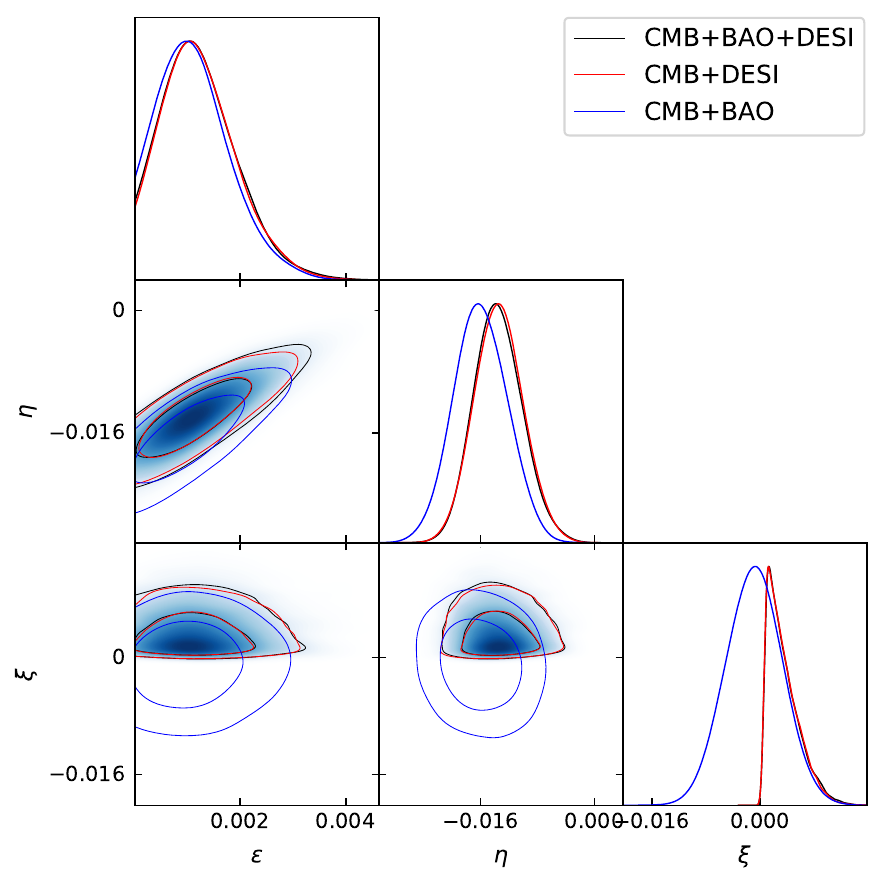}
\caption{The contour plots and likelihood distributions of the slow-roll parameters $\{\epsilon,\eta,\xi\}$ are shown at the $68\%$ and $95\%$ confidence levels for the CMB+BAO+DESI, CMB+DESI, and CMB+BAO data sets, respectively.}
\label{figure2}
\end{figure}

\begin{figure}
\centering
\includegraphics[width=14cm]{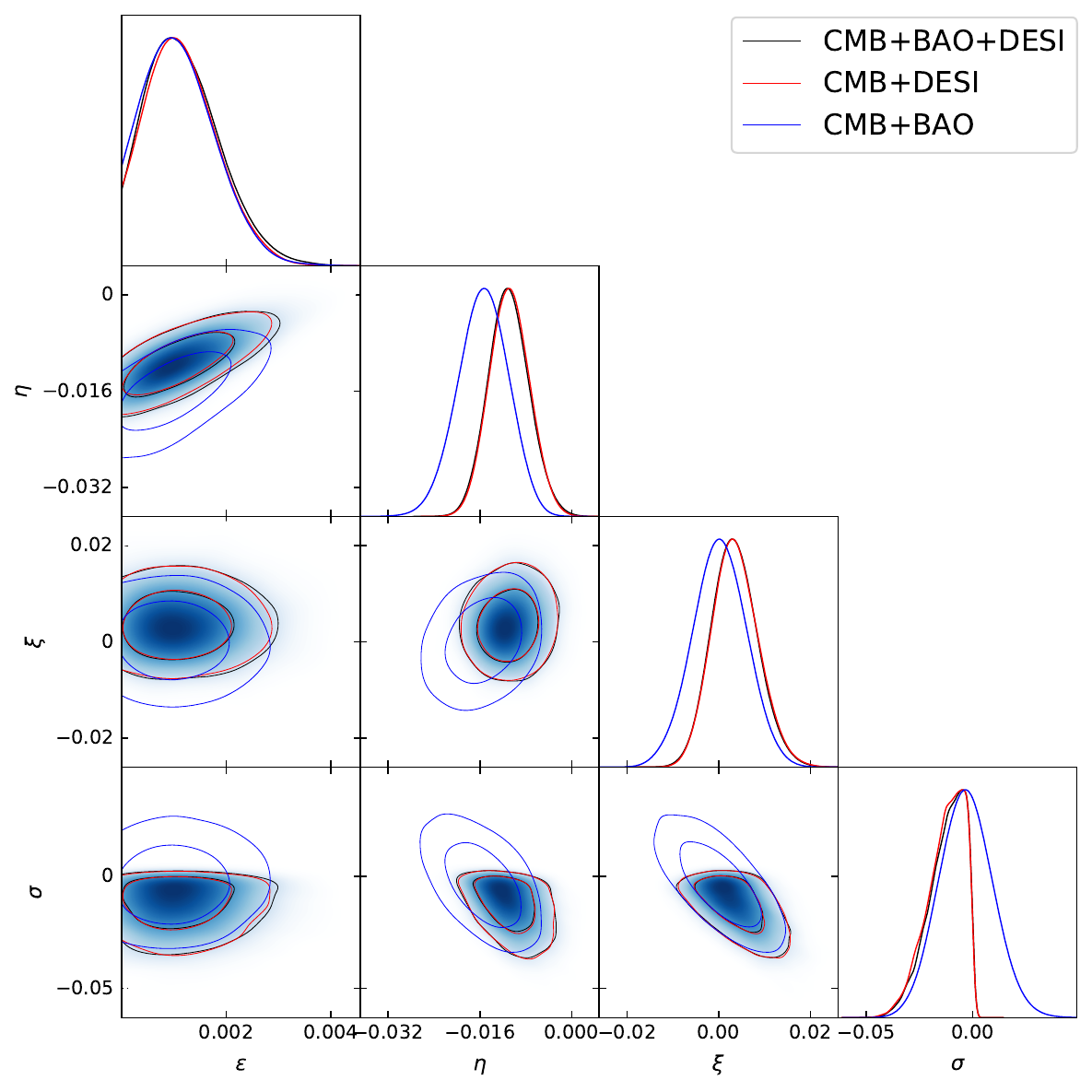}
\caption{The contour plots and likelihood distributions of the slow-roll parameters $\{\epsilon,\eta,\xi, \sigma\}$ are shown at the $68\%$ and $95\%$ confidence levels, based on the CMB+BAO+DESI, CMB+DESI, and CMB+BAO data combinations, respectively.}
\label{figure3}
\end{figure}

In the standard $\Lambda$CDM model, the six parameters are the baryon density parameter $\Omega_b h^2$, the cold dark matter density $\Omega_c h^2$, the angular size of the horizon at the last scattering surface $\theta_\text{MC}$, the optical depth $\tau$, the scalar amplitude $A_s$, and the scalar spectral index $n_s$. This framework can be extended by including the running of the scalar spectral index $\alpha_s$, the running of the running $\beta_s$, and the tensor amplitude $A_t$, or equivalently the tensor-to-scalar ratio $r$. In the present analysis, we replace the usual scalar power-spectrum parameters in the $\Lambda$CDM+$r$ model, the $\Lambda$CDM+$r+\alpha_s$ model, and the $\Lambda$CDM+$r+\alpha_s+\beta_s$ model with the corresponding slow-roll parameters. The parameter sets sampled in CosmoMC are: $\{\Omega_b h^2, \Omega_c h^2, \theta_\text{MC}, \tau, A_s, \epsilon, \eta\}$, $\{\Omega_b h^2, \Omega_c h^2, \theta_\text{MC}, \tau, A_s, \epsilon, \eta, \xi\}$, and $\{\Omega_b h^2, \Omega_c h^2, \theta_\text{MC}, \tau, A_s, \epsilon, \eta, \xi, \sigma\}$. The sampled parameters are assigned uniform priors as follows: $\Omega_bh^2\in[0.005, 0.1]$, $\Omega_ch^2\in[0.001, 0.99]$, $100\theta_{\mathrm{MC}}\in[0.5, 10]$, $\tau\in[0.01, 0.8]$, $\ln\(10^{10}A_s\)\in[2.0, 4.0]$, $\epsilon\in[0, 0.01]$, $\eta\in[-0.1, 0.1]$, $\xi\in[-0.1, 0.1]$, and $\sigma\in[-0.1, 0.1]$. These parameters are then constrained using three combinations of datasets: CMB+BAO+DESI, CMB+DESI, and CMB+BAO. The likelihood for the combined CMB+BAO+DESI dataset includes Planck TTTEEE+lowE+lensing \cite{Planck:2018vyg}, BICEP/Keck 2018 (BK18) \cite{BICEP:2021xfz}, 6dF Galaxy Survey \cite{Beutler:2011hx}, MGS \cite{Ross:2014qpa}, SDSS DR12 \cite{BOSS:2016wmc}, and DESI Data Release 2 (DESI DR2). The specific DESI DR2 data used in this analysis are listed in Table IV of Ref. \cite{DESI:2025zgx}. The resulting constraints on the slow-roll parameters $\{\epsilon, \eta, \xi, \sigma\}$ and the corresponding contour plots are presented in Tables \ref{table1} to \ref{table3} and Figures \ref{figure1} to \ref{figure3}.

\begin{table*}
\newcommand{\tabincell}[2]{\begin{tabular}{@{}#1@{}}#2\end{tabular}}
  \centering
  \begin{tabular}{  c |c| c| c}
  \hline
  \hline
  Parameter & \tabincell{c} {$+\epsilon+\eta$} & \tabincell{c}{$+\epsilon+\eta+\xi$} & \tabincell{c}{$+\epsilon+\eta+\xi+\sigma$}\\
  \hline
  $\Omega_bh^2$ & $0.02247\pm0.00012$   &$0.02258\pm0.00013$ & $0.02254\pm0.00014$ \\
  $\Omega_ch^2$ &$0.11849\pm{0.0005}$    &$0.11785^{+0.0005}_{-0.0006}$ & $0.11779\pm{0.0006}$\\
  $100\theta_{\mathrm{MC}}$ & $1.04117\pm0.00027$   &$1.04123\pm0.00027$ & $1.04121\pm0.00028$\\
  $\tau$ &  $0.0529^{+0.0060}_{-0.0061}$   &$0.0636^{+0.0072}_{-0.0082}$  & $0.0573^{+0.0075}_{-0.0087}$\\
  $\ln\(10^{10}A_s\)$  & $3.040\pm0.012$     &$3.061\pm0.015$ & $3.047^{+0.015}_{-0.017}$\\
  $\epsilon$  & $<0.0025$ & $<0.0025$ & $<0.0025$\\
  $\eta$  & $-0.020^{+0.0061}_{-0.0057}$ & $-0.014^{+0.0071}_{-0.0064}$ & $-0.011^{+0.0072}_{-0.0067}$\\
  $\xi$   & $...$ & $0.0035^{+0.0051}_{-0.0038}$& $0.0035^{+0.0103}_{-0.0093}$\\
  $\sigma$  & $...$ & $...$& $-0.0123^{+0.0147}_{-0.0178}$ \\
  \hline
  \end{tabular}
  \caption{The $68\%$ confidence limits on the cosmological parameters and the $95\%$ confidence limits on the slow-roll parameters are derived from the combined CMB+BAO+DESI datasets.}
  \label{table1}
\end{table*}

\begin{table*}
\newcommand{\tabincell}[2]{\begin{tabular}{@{}#1@{}}#2\end{tabular}}
  \centering
  \begin{tabular}{  c |c| c| c}
  \hline
  \hline
  Parameter & \tabincell{c} {$+\epsilon+\eta$} & \tabincell{c}{$+\epsilon+\eta+\xi$} & \tabincell{c}{$+\epsilon+\eta+\xi+\sigma$}\\
  \hline
  $\Omega_bh^2$ & $0.02247\pm0.00012$   &$0.02258\pm0.00013$ & $0.02255\pm0.00013$ \\
  $\Omega_ch^2$ &$0.11853\pm{0.0006}$    &$0.11779\pm0.0006$ & $0.11768\pm{0.0006}$\\
  $100\theta_{\mathrm{MC}}$ & $1.04116\pm0.00028$   &$1.04124\pm0.00028$ & $1.04121^{+0.00028}_{-0.00027}$\\
  $\tau$ &  $0.0530^{+0.0061}_{-0.0060}$   &$0.0635^{+0.0074}_{-0.0084}$  & $0.0575^{+0.0076}_{-0.0086}$\\
  $\ln\(10^{10}A_s\)$  & $3.040^{+0.013}_{-0.012}$     &$3.060^{+0.015}_{-0.017}$ & $3.048^{+0.015}_{-0.017}$\\
  $\epsilon$  & $<0.0024$ & $<0.0025$ & $<0.0024$\\
  $\eta$  & $-0.020^{+0.0058}_{-0.0055}$ & $-0.013^{+0.0069}_{-0.0065}$ & $-0.011^{+0.0069}_{-0.0066}$\\
  $\xi$   & $...$ & $0.0034^{+0.0049}_{-0.0037}$& $0.0037^{+0.0103}_{-0.0100}$\\
  $\sigma$  & $...$ & $...$& $-0.0126^{+0.0149}_{-0.0178}$ \\
  \hline
  \end{tabular}
  \caption{The $68\%$ confidence limits on the cosmological parameters and the $95\%$ confidence limits on the slow-roll parameters are obtained from the combined CMB+DESI datasets.}
  \label{table2}
\end{table*}

\begin{table*}
\newcommand{\tabincell}[2]{\begin{tabular}{@{}#1@{}}#2\end{tabular}}
  \centering
  \begin{tabular}{  c |c| c| c}
  \hline
  \hline
  Parameter & \tabincell{c} {$+\epsilon+\eta$} & \tabincell{c}{$+\epsilon+\eta+\xi$} & \tabincell{c}{$+\epsilon+\eta+\xi+\sigma$}\\
  \hline
  $\Omega_bh^2$ & $0.02240\pm0.00013$   &$0.02240\pm0.00015$ & $0.02240^{+0.00015}_{-0.00014}$ \\
  $\Omega_ch^2$ &$0.11958\pm{0.0009}$    &$0.11955\pm{0.0009}$ & $0.11954\pm{0.0009}$\\
  $100\theta_{\mathrm{MC}}$ & $1.04099\pm0.00029$   &$1.04099^{+0.00030}_{-0.00029}$ & $1.04098\pm0.00029$\\
  $\tau$ &  $0.0566^{+0.0072}_{-0.0073}$   &$0.0561^{+0.0071}_{-0.0079}$  & $0.0549^{+0.0078}_{-0.0088}$\\
  $\ln\(10^{10}A_s\)$  & $3.048^{+0.014}_{-0.015}$     &$3.047^{+0.014}_{-0.016}$ & $3.045^{+0.016}_{-0.017}$\\
  $\epsilon$  & $<0.0023$ & $<0.0024$ & $<0.0023$\\
  $\eta$  & $-0.016^{+0.0072}_{-0.0070}$ & $-0.016^{+0.0075}_{-0.0073}$ & $-0.015^{+0.0082}_{-0.0088}$\\
  $\xi$   & $...$ & $-0.0009\pm{0.0082}$& $0.0002^{+0.0114}_{-0.0116}$\\
  $\sigma$  & $...$ & $...$& $-0.0031^{+0.0252}_{-0.0242}$ \\
  \hline
  \end{tabular}
  \caption{The $68\%$ confidence limits on the cosmological parameters and the $95\%$ confidence limits on the slow-roll parameters are obtained from the combined CMB+BAO datasets.}
  \label{table3}
\end{table*}

From the results, we observe that the constraint on the first slow-roll parameter $\epsilon$ varies only slightly when other slow-roll parameters are included in the analysis. The central value of the second slow-roll parameter $\eta$ shifts modestly upon introducing the third and fourth parameters $\xi$ and $\sigma$, while its uncertainty range widens noticeably. A similar trend is seen for the third slow-roll parameter $\xi$: its central value changes little when $\sigma$ is added, but the allowed range between the upper and lower limits becomes broader.

Compared with earlier data combinations, the inclusion of DESI Data Release 2 has a pronounced impact on the constraints on $\eta$, $\xi$, and $\sigma$, whereas the constraint on $\epsilon$ remains largely unchanged. BAO distance measurements mainly constrain the late-time expansion history and only indirectly affect the inflationary parameters. Since DESI data do not directly constrain the slow-roll parameters, these shifts may arise from several sources: correlations between the CMB-inferred scalar tilt and BAO-related background parameters, possibly linked to the BAO--CMB tension \cite{Ferreira:2025lrd}; or correlated shifts involving the optical depth, together with the effects of CMB lensing and large-scale polarization choices \cite{McDonough:2025lzo}---a question that remains open.

\section{constraints on the power spectra parameters from observations}
We then derive the slow-roll inflation predictions consistent with the observational constraints. The parameters $\{r, n_t, \alpha_t, n_s, \alpha_s, \beta_s\}$, which characterize the scalar and tensor power spectra, are evaluated using Eqs.~(\ref{slr}) - (\ref{slbetas}). These quantities are treated as derived parameters. Sampling $\{\epsilon, \eta, \xi, \sigma\}$ induces nontrivial priors on the power spectrum parameters. Specifically, the induced priors are bounded as follows: $r\leq0.187$, $|n_t|\leq 2.04\times 10^{-2}$, $|\alpha_t|\leq4.66\times 10^{-3}$, $n_s\in[0.762, 1.207]$, $\alpha_s\in[-0.241, 0.241]$, and $\beta_s\in[-0.294, 0.356]$. The resulting values are presented in Tables \ref{table4} to \ref{table6} and Figures \ref{figure4} to \ref{figure9}.

From the results, we observe that the inclusion of DESI DR2 does not improve the upper bound on $r$. According to Eq.~(\ref{slr}), the tensor-to-scalar ratio $r$ is determined by the slow-roll parameters $\epsilon$ and $\eta$. Since DESI DR2 does not tighten the upper bound on $\epsilon$, and the shift in the central value of $\eta$ is only about 0.004, which is negligible compared with unity, the inclusion of DESI DR2 consequently has no significant effect on the upper bound of $r$.
In what follows, we analyse the scalar and tensor spectral parameters separately.

\begin{table*}
\newcommand{\tabincell}[2]{\begin{tabular}{@{}#1@{}}#2\end{tabular}}
  \centering
  \begin{tabular}{  c |c| c| c}
  \hline
  \hline
  Parameter & \tabincell{c} {$\Lambda$CDM+$r$} & \tabincell{c}{$\Lambda$CDM+$r+\alpha_s$} & \tabincell{c}{$\Lambda$CDM+$r+\alpha_s+\beta_s$}\\
  \hline
  $r$ &  $<0.038$   &$<0.039$  & $<0.038$\\
  $-n_t(\times10^{-2})$  & $<0.49$     &$<0.50$ & $<0.49$\\
  $-\alpha_t(\times10^{-4})$  & $<1.91$ & $<1.44$&$<1.22$ \\
    $n_s$ & $0.9539^{+0.0027}_{-0.0023}$   &$0.9657^{+0.0047}_{-0.0042}$ & $0.9710\pm0.0051$ \\
  $\alpha_s$ &$...$    &$-0.0069^{+0.0059}_{-0.0023}$ & $-0.0070^{+0.0104}_{-0.0092}$\\
  $\beta_s$ & $...$   &$...$ & $-0.0230^{+0.0218}_{-0.0102}$\\
  \hline
  \end{tabular}
  \caption{The $95\%$ confidence limits on the tensor power spectrum parameters $r$, $n_t$, and $\alpha_t$, as well as the $68\%$ confidence limits on the scalar power spectrum parameters $n_s$, $\alpha_s$, and $\beta_s$, are derived from combinations of the CMB+BAO+DESI datasets.}
  \label{table4}
\end{table*}

\begin{table*}
\newcommand{\tabincell}[2]{\begin{tabular}{@{}#1@{}}#2\end{tabular}}
  \centering
  \begin{tabular}{  c |c| c| c}
  \hline
  \hline
  Parameter & \tabincell{c} {$\Lambda$CDM+$r$} & \tabincell{c}{$\Lambda$CDM+$r+\alpha_s$} & \tabincell{c}{$\Lambda$CDM+$r+\alpha_s+\beta_s$}\\
  \hline
  $r$ &  $<0.037$   &$<0.039$  & $<0.037$\\
  $-n_t(\times10^{-2})$  & $<0.48$     &$<0.50$ & $<0.47$\\
  $-\alpha_t(\times10^{-4})$  & $<1.88$ & $<1.43$&$<1.15$ \\
    $n_s$ & $0.9538^{+0.0026}_{-0.0022}$   &$0.9659^{+0.0049}_{-0.0042}$ & $0.9714\pm0.0051$ \\
  $\alpha_s$ &$...$    &$-0.0068^{+0.0058}_{-0.0023}$ & $-0.0073^{+0.0106}_{-0.0092}$\\
  $\beta_s$ & $...$   &$...$ & $-0.0237^{+0.0222}_{-0.0104}$\\
  \hline
  \end{tabular}
  \caption{The $95\%$ confidence intervals for the tensor power spectrum parameters $r$, $n_t$, and $\alpha_t$, as well as the $68\%$ confidence intervals for the scalar power spectrum parameters $n_s$, $\alpha_s$, and $\beta_s$, are derived from the combined CMB+DESI datasets.}
  \label{table5}
\end{table*}

\begin{table*}
\newcommand{\tabincell}[2]{\begin{tabular}{@{}#1@{}}#2\end{tabular}}
  \centering
  \begin{tabular}{  c |c| c| c}
  \hline
  \hline
  Parameter & \tabincell{c} {$\Lambda$CDM+$r$} & \tabincell{c}{$\Lambda$CDM+$r+\alpha_s$} & \tabincell{c}{$\Lambda$CDM+$r+\alpha_s+\beta_s$}\\
  \hline
  $r$ &  $<0.036$   &$<0.037$  & $<0.036$\\
  $-n_t(\times10^{-2})$  & $<0.47$     &$<0.48$ & $<0.47$\\
  $-\alpha_t(\times10^{-4})$  & $<1.56$ & $<1.57$&$<1.56$ \\
    $n_s$ & $0.9613^{+0.0056}_{-0.0050}$   &$0.9613^{+0.0058}_{-0.0052}$ & $0.9625^{+0.0079}_{-0.0068}$ \\
  $\alpha_s$ &$...$    &$0.0014^{+0.0080}_{-0.0079}$ & $-0.0007\pm0.0111$\\
  $\beta_s$ & $...$   &$...$ & $-0.0059^{+0.0231}_{-0.0233}$\\
  \hline
  \end{tabular}
  \caption{The $95\%$ confidence limits on the tensor power spectrum parameters $r$, $n_t$, and $\alpha_t$, as well as the $68\%$ confidence limits on the scalar power spectrum parameters $n_s$, $\alpha_s$, and $\beta_s$, are derived from the combined CMB+BAO datasets.}
  \label{table6}
\end{table*}

\subsection{constraints on the scalar power spectrum parameters from observations}

\begin{figure}
\centering
\includegraphics[width=9cm]{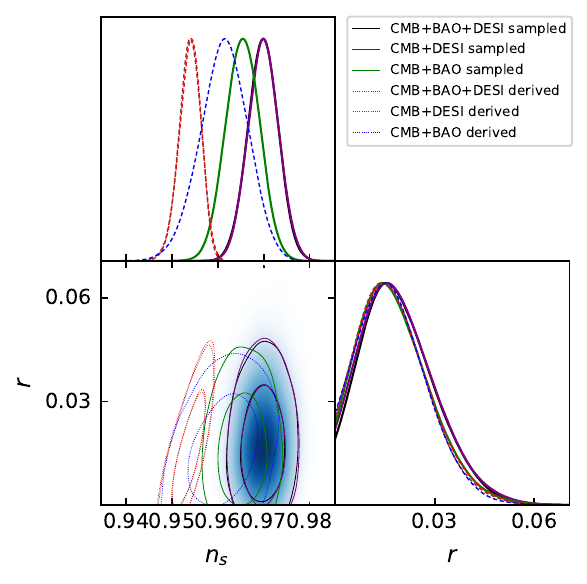}
\caption{The contour plots and likelihood distributions of the scalar power spectrum parameters $\{r, n_s\}$ are shown at the $68\%$ and $95\%$ confidence levels for the CMB+BAO+DESI, CMB+DESI, and CMB+BAO data combinations, respectively. The filled lines represent the sampled results, whereas the dashed lines represent the derived results.}
\label{figure4}
\end{figure}

\begin{figure}
\centering
\includegraphics[width=11cm]{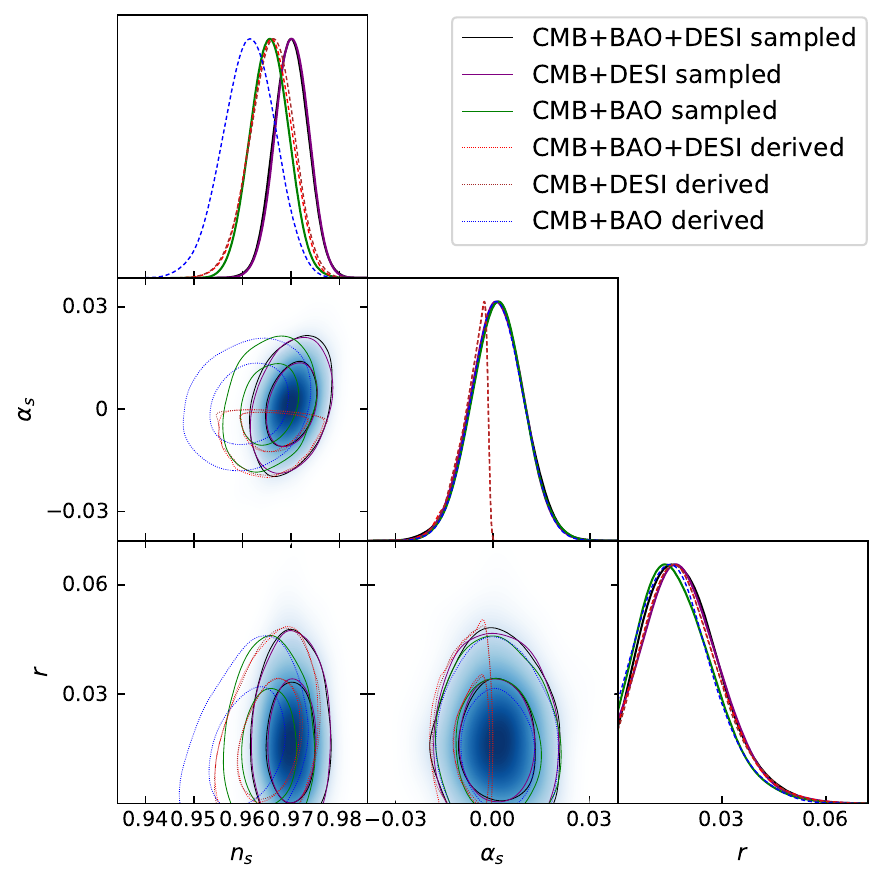}
\caption{The contour plots and likelihood distributions of the scalar power spectrum parameters $\{r, n_s, \alpha_s\}$ are shown at the $68\%$ and $95\%$ confidence levels, based on the CMB+BAO+DESI, CMB+DESI, and CMB+BAO dataset combinations, respectively. The filled lines represent the sampled results, and the dashed lines represent the derived results.}
\label{figure5}
\end{figure}

\begin{figure}
\centering
\includegraphics[width=13cm]{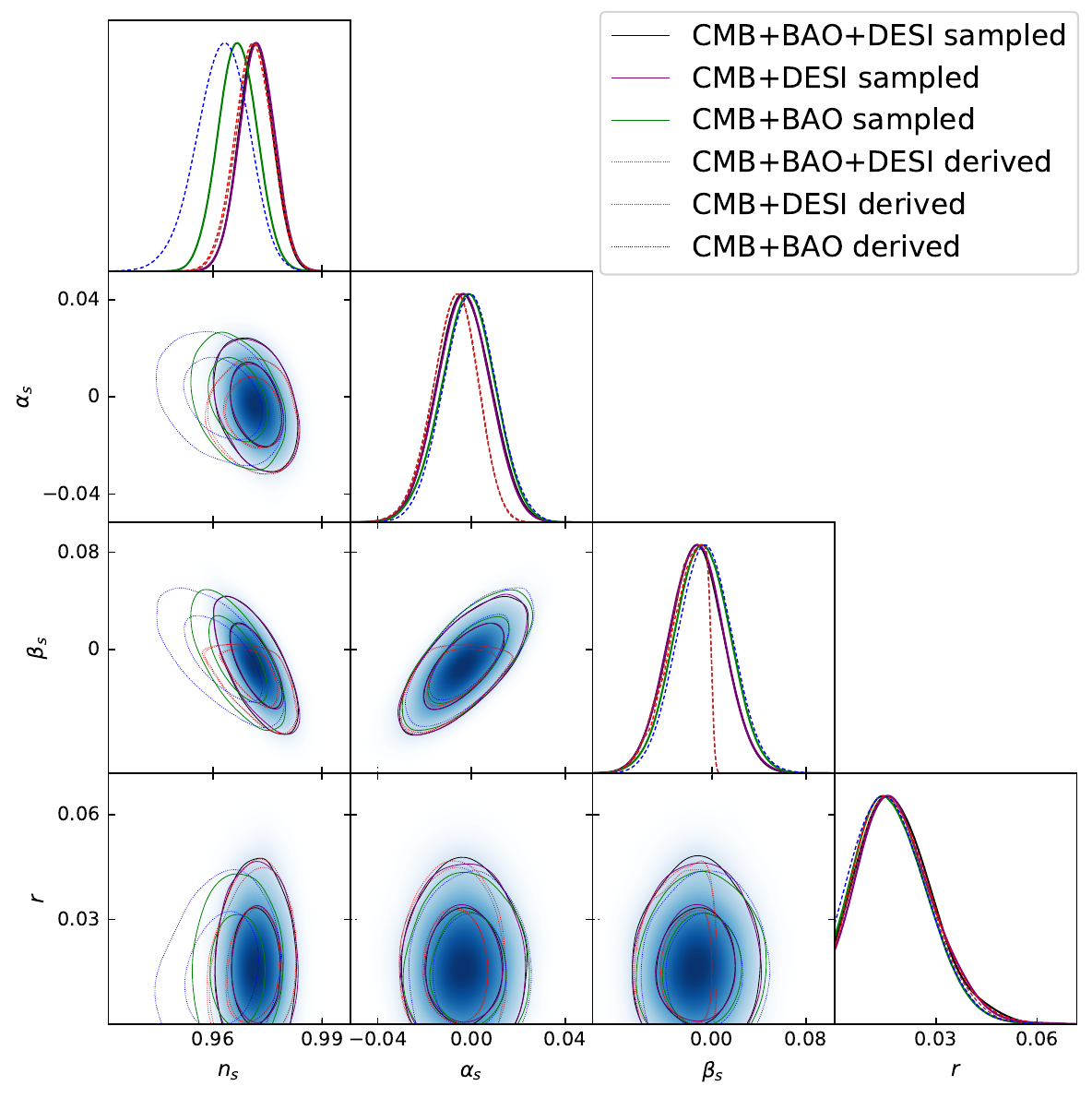}
\caption{The contour plots and likelihood distributions of the scalar power-spectrum parameters $\{r, n_s, \alpha_s, \beta_s\}$ are shown at the $68\%$ and $95\%$ confidence levels for the CMB+BAO+DESI, CMB+DESI, and CMB+BAO data combinations, respectively. The solid lines represent the sampled results, and the dashed lines represent the derived results.}
\label{figure6}
\end{figure}

In this subsection, we focus on the derived scalar power spectrum parameters: the spectral index $n_s$, its running $\alpha_s$, and the running of the running $\beta_s$. Tables \ref{table4} to \ref{table6} show that the central values of $n_s$ shift only modestly when $\alpha_s$ and $\beta_s$ are treated as free parameters, whereas the uncertainties, quantified by the difference between the upper and lower bounds, increase noticeably. A similar trend is observed for $\alpha_s$: its central value changes little when $\beta_s$ is added, but the allowed range widens substantially. In our earlier study \cite{Li:2022avp}, the parameters $n_s$, $\alpha_s$, and $\beta_s$ were treated as directly sampled parameters. In the present work, we extend that analysis by considering two additional dataset combinations, CMB+BAO+DESI and CMB+DESI, alongside the original CMB+BAO baseline. The resulting constraints are shown in Figures \ref{figure4} to \ref{figure6}, allowing a direct comparison between the derived values obtained here through the slow-roll mapping and the sampled values. The derived parameters remain broadly consistent with the sampled ones, except for $n_s$. For the spectral index, the derived values are systematically lower than the sampled values, a shift that is clearly visible in Figures \ref{figure4} to \ref{figure6}. This discrepancy suggests that the slow-roll relations impose a tighter correlation between $n_s$ and the other inflationary parameters, shifting its preferred range toward slightly smaller values.

It is therefore necessary to investigate whether the differences between the derived and sampled results originate from the likelihood, the slow-roll consistency relations, prior-volume effects, or the choice of prior boundaries. To isolate the source of this discrepancy, we consider three dataset combinations: CMB+BAO+DESI, CMB+DESI, and CMB+BAO. Within each combination, the derived and sampled analyses share the same underlying data sets and therefore yield the same likelihood. Hence, the likelihood itself cannot account for the observed differences within a given combination. However, across different combinations, the likelihood does produce noticeable effects on the results. In the derived analysis, we employ the slow-roll consistency relations given by Eqs.~(\ref{slr})--(\ref{slbetas}). In contrast, the sampled analysis adopts the leading-order consistency relation of single-field slow-roll inflation, namely $n_t=-r/8$ \cite{Liddle:1992wi, Copeland:1993ie}, together with $\alpha_t=0$. The discrepancies observed between the derived and sampled results are attributable to the different choices of consistency relations. Prior-volume effects refer to the phenomenon in which the posterior distribution becomes non-uniformly weighted due to differences in the volume of the parameter space. When mapping from slow-roll parameters to derived parameters via a nonlinear transformation, even if the priors on the slow-roll parameters are uniform, the effective priors on the derived parameters become non-uniform owing to the Jacobian determinant. This effect can cause certain regions of the parameter space to be statistically overrepresented or underrepresented, thereby affecting the final constraints. The prior boundaries also differ between the derived and sampled analyses. Sampling the slow-roll parameters induces nontrivial priors on the derived power-spectrum parameters, although these induced priors are much broader than the resulting constraints.

\begin{figure}
\centering
\includegraphics[width=8cm]{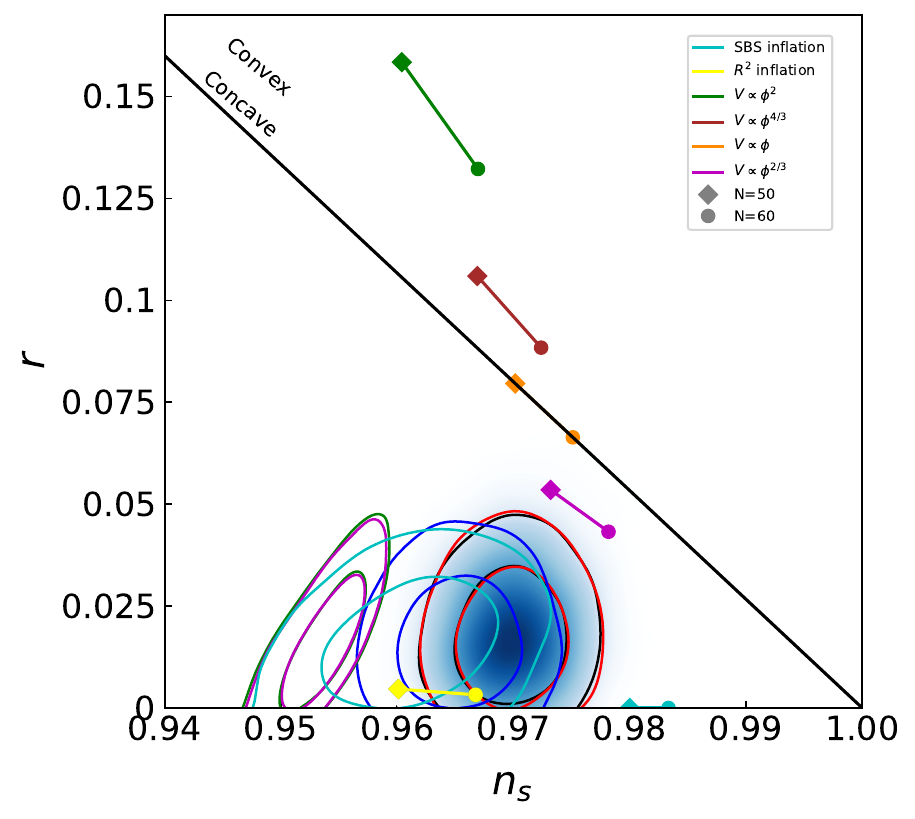}
\caption{The contour plot of the scalar power-spectrum parameters in the $n_s$-$r$ plane is shown at the $68\%$ and $95\%$ confidence levels, derived from the CMB+BAO+DESI, CMB+DESI, and CMB+BAO data combinations, respectively. The curved black, red, and blue lines represent the sampled results, whereas the curved green, purple, and cyan lines represent the derived results.}
\label{figure7}
\end{figure}

We then compare the constraints in the $r-n_s$ plane from cosmological data sets with the predictions of several simple inflationary models. The systematic difference between the derived and sampled values of $n_s$ may arise from the different slow-roll consistency relations, prior-volume effects, or the choice of prior boundaries. Our main results are summarised in Fig.~\ref{figure7}, which shows that inflationary models with a concave potential are favoured at more than $95\%$ confidence level. A detailed discussion follows.

Among the simplest realisations of inflation are models with a monomial potential $V(\phi)\sim \phi^n$ \cite{Linde:1983gd}. In the slow-roll approximation, their predictions are
\m
r&\simeq&\frac{16n}{4N+n}, \\
n_s&\simeq&1-\frac{2(n+2)}{4N+n}.
\n
Here $N$ denotes the number of e-folds before the end of inflation, and $n$ need not be an integer. For instance, axion-monodromy constructions in string theory can yield potentials with $n=2/5$, $2/3$ \cite{Silverstein:2008sg}, $n=1$ \cite{McAllister:2008hb}, or even higher powers \cite{Marchesano:2014mla,McAllister:2014mpa}. The predicted trajectories for $N\in [50,60]$ are shown in Fig.~\ref{figure7} as lines connecting diamond symbols (for $N=50$) to circles (for $N=60$). The entire family lies outside the $95\%$ CL contour obtained from the data, indicating that monomial-potential inflation is disfavoured at this confidence level.

The spontaneously broken supersymmetry (SBS) inflation model \cite{Dvali:1994ms,Copeland:1994vg,Binetruy:1996xj,Stewart:1994ts,Lyth:1998xn} is characterized by a potential of the form $V(\phi)=V_0\left(1+c\ln\frac{\phi}{Q}\right)$, where $V_0$ dominates and $c<<1$. In this model, the predicted tensor-to-scalar ratio and scalar spectral index are
\m
r&\simeq&0, \\
n_s&=&1-\frac{1}{N}.
\n
The corresponding point in the $r-n_s$ plane lies outside the $95\%$ CL contour from current data, so the basic SBS model is disfavoured at this confidence level. It has been noted, however, that including soft SUSY-breaking terms can shift $n_s$ sufficiently to bring the predictions into agreement with observations \cite{Rehman:2009nq}.

The Starobinsky inflation model \cite{Starobinsky:1980te} arises from a gravitational action containing a quadratic Ricci scalar term: $S=\frac{M_p^2}{2}\int d^4x\sqrt{-g}\left(R+\frac{R^2}{6M^2}\right)$, where $M$ is a mass scale. In the slow-roll approximation, the model predicts
\m
r&\simeq&\frac{12}{N^2}, \\
n_s&=&1-\frac{2}{N},
\n
as derived in \cite{Mukhanov:1981xt,Starobinsky:1983zz}. The Starobinsky predictions fall well within the observationally allowed region and are strongly favoured by the data.

The detailed physics of reheating after inflation remains uncertain, leading to an ambiguity in the exact number of e-folds $N$ between the time when the pivot scale $k_*$ left the horizon and the end of inflation. This uncertainty translates into a corresponding uncertainty in the precise predictions of inflationary models. To circumvent this issue, we follow the approach introduced in \cite{Huang:2007qz,Huang:2015cke,Li:2019efi} and parameterize the first slow-roll parameter as a function of $N$:
\m
\epsilon=\frac{q/2}{\left(N+\Delta N\right)^p},\quad \Delta N=\left(\frac{q}{2}\right)^{1/p},
\n
where $q$ and $p$ are dimensionless constants. Within this parametrization, the tensor-to-scalar ratio and the scalar spectral index become
\m
r&=&\frac{8q}{\left(N+\Delta N\right)^p},\\
n_s&=&1-\frac{q}{\left(N+\Delta N\right)^p}-\frac{p}{N+\Delta N}.
\n
This form is flexible enough to reproduce a wide range of well-studied inflationary potentials. For example, $p=1$ and $q=n/2$ for $V(\phi)\sim \phi^n$, $p=2$ and $q=3/2$ for the Starobinsky inflation model, and $p=2(d-1)/d$ and $q\simeq 0$ for the brane inflation model \cite{Dvali:1998pa,Kachru:2003sx} with potential $V(\phi)=V_0(1-(\mu/\phi)^{d-2})$.

\begin{figure}
\centering
\includegraphics[width=8cm]{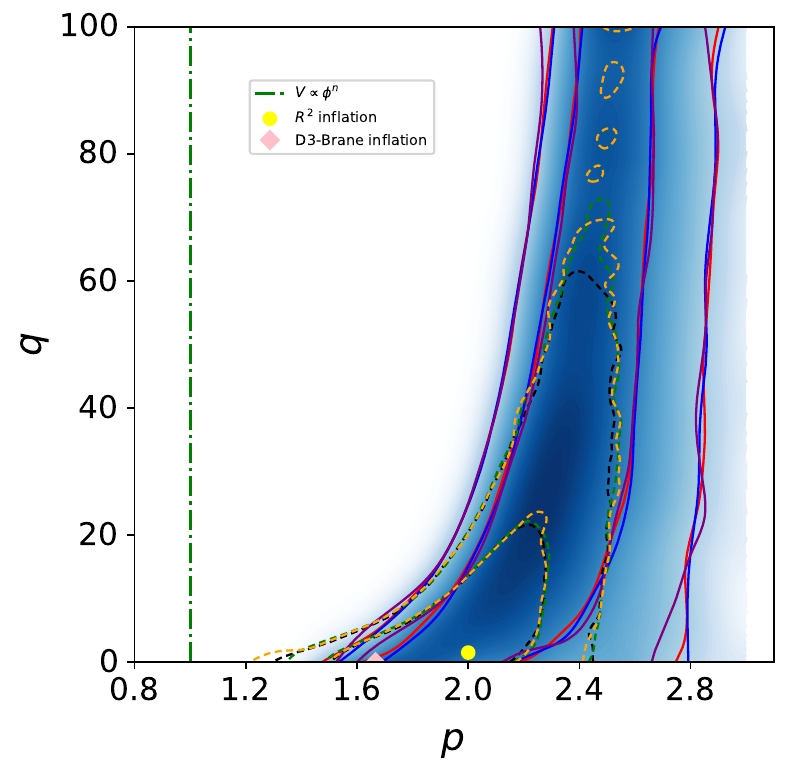}
\caption{The contour plot of $p-q$ is shown at the 68\% and 95\% confidence levels for the CMB+BAO+DESI, CMB+DESI, and CMB+BAO datasets, respectively. The curved dashed lines represent the results for $N\in [50,60]$. The curved filled lines represent the results for $N\in [14,75]$.}
\label{figure8}
\end{figure}

\begin{figure}
\centering
\includegraphics[width=11cm]{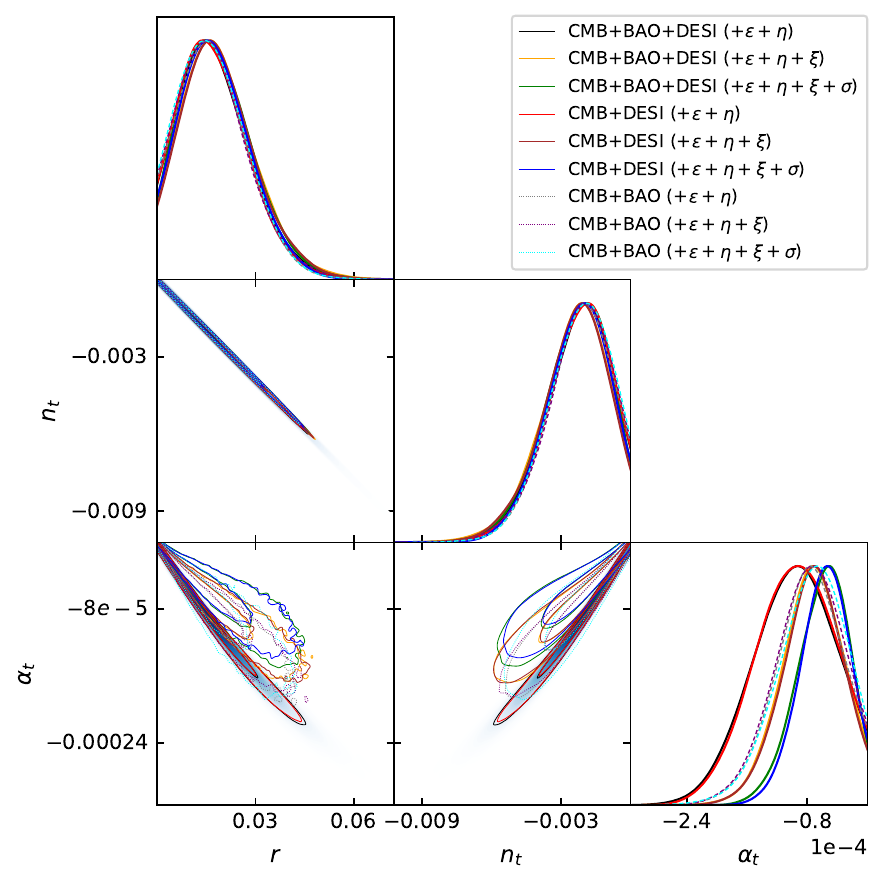}
\caption{The contour plots and likelihood distributions of the tensor power spectrum parameters are shown at the $68\%$ and $95\%$ confidence levels for the CMB+BAO+DESI, CMB+DESI, and CMB+BAO datasets, respectively.}
\label{figure9}
\end{figure}

In our analysis, we treat $N$, $p$ and $q$ as free parameters, thereby marginalising over the reheating uncertainty. Adopting the conventional range $N\in [50,60]$, the CMB+BAO+DESI dataset gives
\m
p &=&  2.07^{+0.24}_{-0.23}\quad(68\% \ \mathrm{CL}),\\
q &<&  13.1\quad(68\% \ \mathrm{CL}).
\n
If we instead adopt a more conservative e-fold range, $N\in [14,75]$ as suggested in \cite{Alabidi:2006qa}, the constraints loosen to
\m
p &=&  2.40^{+0.27}_{-0.23}\quad(68\% \ \mathrm{CL}),\\
q &<&  58.9\quad(68\% \ \mathrm{CL}).
\n
In both cases, the models with $p=1$ corresponding to $V(\phi)\sim \phi^n$ are excluded at more than $95\%$ CL. In contrast, the Starobinsky model and brane inflation model remain fully consistent with the data. These results are displayed graphically in Fig.~\ref{figure8}. The constraints on the parameters $p$ and $q$ provide valuable insight into both the reheating uncertainty and the classification of representative inflationary models. By treating the number of e-folds $N$ as a free parameter and marginalizing over it, we effectively incorporate the unknown reheating physics into the analysis, allowing the data to directly inform the allowed range of inflationary predictions. The $(p,q)$ parametrization not only marginalizes over the reheating uncertainty in a principled manner, but also provides a robust framework for classifying and testing broad families of inflationary models against observational constraints.

\subsection{constraints on the tensor power spectrum parameters from observations}
In this subsection, we analyze the derived parameters of the tensor power spectrum: the tensor-to-scalar ratio $r$, the tensor spectral index $n_t$, and its running $\alpha_t$. Within the constrained single-field slow-roll framework, both $n_t$ and its running $\alpha_t$ are predicted to be negative. The derived tensor spectral index is very small, with $|n_t|\lesssim 4.9\times 10^{-3}$ at the $95\%$ confidence level. A value of this magnitude will be extremely challenging to detect with CMB observations alone in the foreseeable future. Similarly, the absolute value of the derived running of the tensor spectral index satisfies $|\alpha_t|\lesssim1.91\times 10^{-4}$ from the combination of the CMB+BAO+DESI datasets at the $95\%$ confidence level.

\section{summary}
In this work, we constrain the canonical single-field slow-roll inflationary framework using two complementary approaches. The first method exploits the analytic dependence of primordial perturbations on the slow-roll parameters, while the second adopts a phenomenological parameterization of the scalar and tensor primordial spectra. We derive direct constraints on the slow-roll parameters using the latest cosmological datasets, notably the BICEP/Keck CMB polarization measurements, which currently provide the tightest constraints on the tensor-to-scalar ratio. A key advantage of this strategy is that it allows us to compute the predictions of single-field slow-roll inflation directly from the constrained parameter values.

We present the resulting predictions for the parameters that characterize the scalar power spectrum and use them to constrain representative inflationary models. Our analysis shows that monomial-potential inflation is disfavored, whereas models with a concave potential, such as the Starobinsky model and brane inflation, are preferred. We also derive the corresponding predictions for the tensor power-spectrum parameters, finding that both the tensor spectral index and its running are negative. Their absolute values are bounded as $|n_t|\lesssim 4.9\times 10^{-3}$ and $|\alpha_t|\lesssim1.91\times 10^{-4}$ at $95\%$ confidence level, based on the combination of CMB+BAO+DESI datasets. These results imply that detecting either $n_t$ or $\alpha_t$ with future observations will be extremely challenging.

\noindent{\bf Acknowledgments}.
This work is supported by the National Natural Science Foundation of China (Grant No. 12405069), the Natural Science Foundation of Shandong Province (Grant No. ZR2021QA073), and the Research Start-up Fund of QUST (Grant No. 1203043003587).

\end{document}